# Thermal Inverse design for resistive micro-heaters

Khoi Phuong Dao, and Juejun Hu

Department of Materials Science and Engineering, Massachusetts Institute of Technology, Cambridge, MA 02139, USA

**Abstract**

This paper proposes an inverse design scheme for resistive heaters. By adjusting the spatial distribution of a binary electrical resistivity map, the scheme enables objective-driven optimization of heaters to achieve pre-defined steady-state temperature profiles. The approach can be fully automated and is computationally efficient since it does not entail extensive iterative simulations of the entire heater structure. The design scheme offers a powerful solution for resistive heater device engineering in applications spanning electronics, photonics, and microelectromechanical systems.

## Introduction

Recent years have witnessed a surge of interest in inverse design, which generally refer to algorithms that automate design of physical systems based on pre-defined performance targets. This is in contrast to conventional forward design approaches, where multiple candidates are evaluated to down select an optimized design, either by exhausting the entire design space or through human-guided design iterations. This inverse design paradigm has transformed many fields in engineering, leading to photonic, mechanical, acoustic, and magnetic devices with non-intuitive configurations and/or unprecedented functions such as nanophotonic devices [1–3], metasurface optics, mechanical metamaterials [4], phononic crystals [5], and micro-magnet arrays [6].

In this work, we aim to develop an inverse design strategy for an important class of devices: resistive micro-heaters, whose applications are pervasive in electronics, photonics, microelectromechanical systems (MEMS), and microfluidics. Design of these heaters have been implemented via numerical optimization in a trial-and-error process, where the heater geometry is adjusted iteratively based on finite element method (FEM) simulation results of the temperature map across the heater [7–10]. Alternatively, the general Cauchy-type thermal inverse design problem, which seeks to optimize a heating device (comprising a spatial distribution of heat sources) to produce a pre-defined temperature profile, has also been approached using variational methods and genetic algorithms [11–13]. However, all these methods suffer from one intrinsic limitation: they necessitate modeling the entire heater device in each iteration. Such "global" simulations are computationally expensive, which limits either the domain size or the number of iterations that can be executed. As a result, the aforementioned methods will only work for very small heaters or simple heat source configurations that can be parameterized with a small set of variables. To circumvent this limitation, our approach uses only parameterized thermal properties of unit cells to deterministically infer the heater design, thereby curtailing the need for "global" simulation iterations. Using doped silicon resistive heaters as an example, we show that our objective-driven approach allows automated generation of micro-heater designs with on-demand steady-state temperature profiles.

## Method

We present our inverse design scheme in doped silicon micro-heaters fabricated on a silicon-on-insulator (SOI) platform, although the methodology can be readily generalized to other types of resistive heaters. Doped SOI heaters are fully CMOS-compatible, and can be seamlessly integrated with frontend electronic and photonic devices for applications spanning thermo-optic switching, material processing, gas sensing [14–16], and beyond. Electrical current paths in doped Si heaters can be defined with the doping profile without altering the physical structure, a feature that is particularly useful for photonic devices since the thermal and optical design considerations can be decoupled. These features ensure that our work not only informs a general inverse design approach but also have significant practical relevance.

The inverse design problem is formulated as follows. A steady-state temperature distribution $T(x, y)$, where $x, y$ are the in-plane coordinates on the heater surface, is defined as the design objective. Figure 1 depicts the heater layout under consideration, consisting of a SOI slab connecting two constant-biased electrical contacts. The target temperature profile is specified in a domain encircled by the red solid lines. The design flow starts with defining a mesh to divide the domain into a set of unit cells with the condition that the temperature across a single unit cell is approximated as uniform. The doping regions have a uniform dopant concentration, and they form an array of filaments with varying widths $W(x, y)$, each traversing a continuous row of unit cells connecting the two contacts. This binary doping pattern can be fabricated by lithography followed by ion implantation. As we shall see later, given a heat generation rate profile $g(x, y)$ (defined as the heat generated in each unit cell per unit time in Watts), one can deterministically solve a unique solution $W(x, y)$ for each and every rows of unit cells.

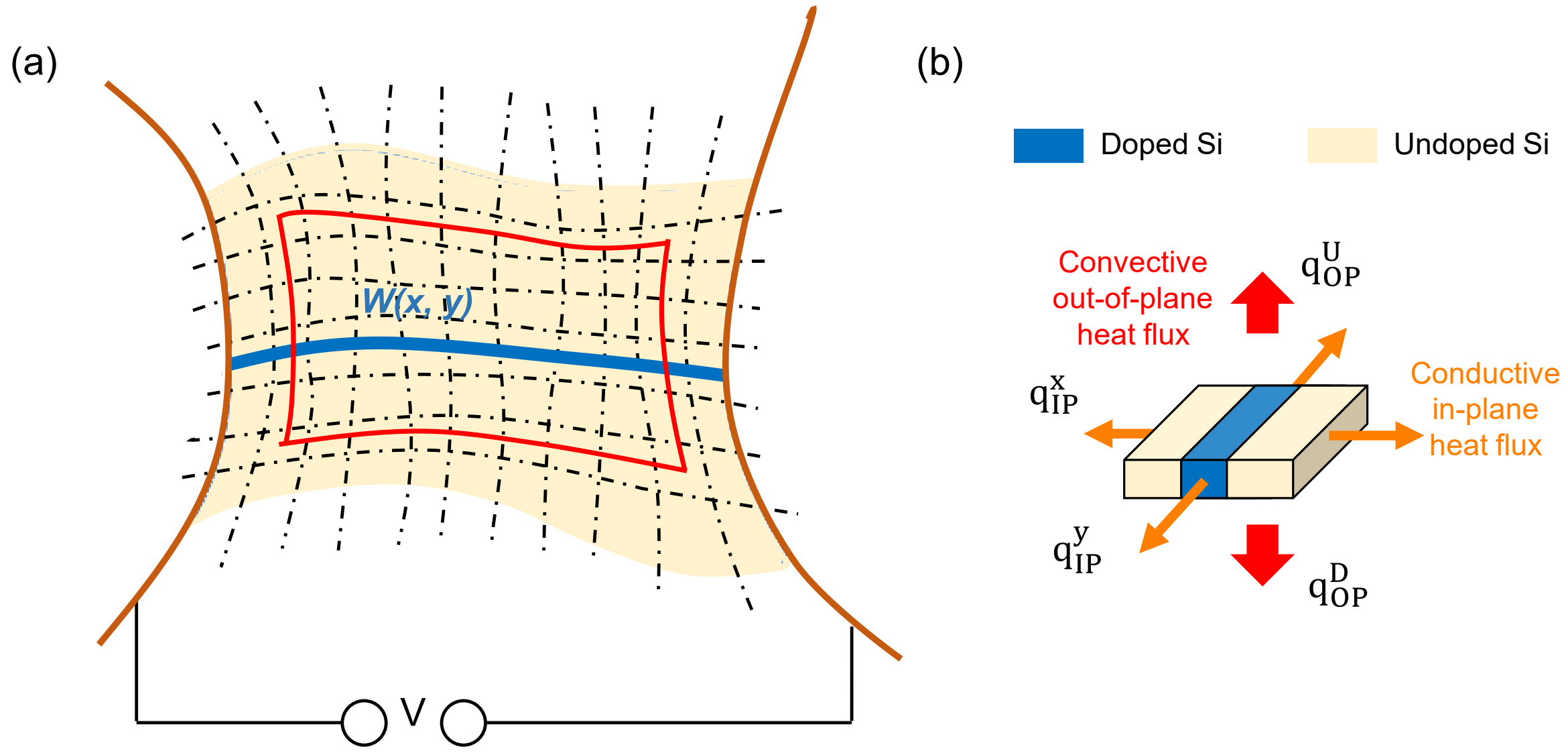

Fig. 1. (a) Schematic depiction of the heater, consisting of a Si slab connecting two constant-biased electrical contacts. The red lines encircle the "target zone" where the target temperature distribution is defined, and the blue strip labels a doped region. (b) A unit cell configuration showing the heat transfer directions.

Next we consider the inverse problem of solving the heat source distribution $g(x, y)$ from $T(x, y)$. We note that the heat fluxes into and out of a unit cell can be determined solely based on its local environment, i.e., the temperatures of the unit cell and its neighboring cells. Specifically, we hypothesize that heat transfer from a unit cell to the outside can be partitioned into out-of-plane and in-plane components, and the transfer rate only depends on the temperatures of the unit cell, its neighboring cells, and the heat sink (the ambient environment). This approximation is valid under the condition that the in-plane temperature gradient is much smaller than the out-of-plane temperature gradient in SOI heaters. The out-of-plane heat transfer rate (in Watts) can be written as:

$$q_{OP} = C_1(T - T_R) \tag{1}$$

where $C_1$ is the product of the equivalent out-of-plane thermal conductance and the unit cell top/bottom surface area, $T_R$ denotes room temperature, and $T$ designates the temperature of the unit cell with the implicit assumption that the temperature within a unit cell is treated as uniform. The in-plane heat transfer rate follows the Fourier's law:

$$q_{IP,x/y} = C_2 \nabla_{x/y} T \tag{2}$$

Similarly, $C_2$ corresponds to the product of the equivalent in-plane thermal conductance and the unit cell side area, and $\nabla_{x/y} T$ denotes the temperature gradient in the in-plane $x$ or $y$ direction. The heat generation $g(x, y)$ can be solved by summing the heat transfer rates over all directions in the steady state.

We then use FEM simulations shown in Figure 2 to validate Eqs. (1) and (2), and derive the constitutive parameters $C_1$ and $C_2$. It is noteworthy that the simulations are not performed over the whole heater, but only across a small domain that contains the target unit cell (marked by red dotted lines) and its neighbors. In our example, the FEM simulation domain contains 5 × 5 square unit cells, each having a size of 1 μm × 1 μm. In the more general case, for an arbitrary heater geometry, the meshing process leads to unit cells of parallelogram shapes. In this case, a pair of $C_1$ and $C_2$ parameters need to be computed for each unique unit cell shape to create a lookup table that can then be repeatedly used for different target temperature profiles. The heater stack comprises (from top to bottom) air, a 10 nm thick silicon dioxide layer, 220 nm SOI layer, 3 μm buried oxide (BOX), and the Si substrate which acts as the heat sink. Stripes of doped Si (labeled with blue-gray color) with a fixed doping concentration of $10^{20}$ $cm^{-3}$ traverse the unit cell columns. The constitutive parameters only weakly depend on the doping region width, and therefore we only perform the simulation for two widths and use linear interpolation to estimate the parameters for other width values.

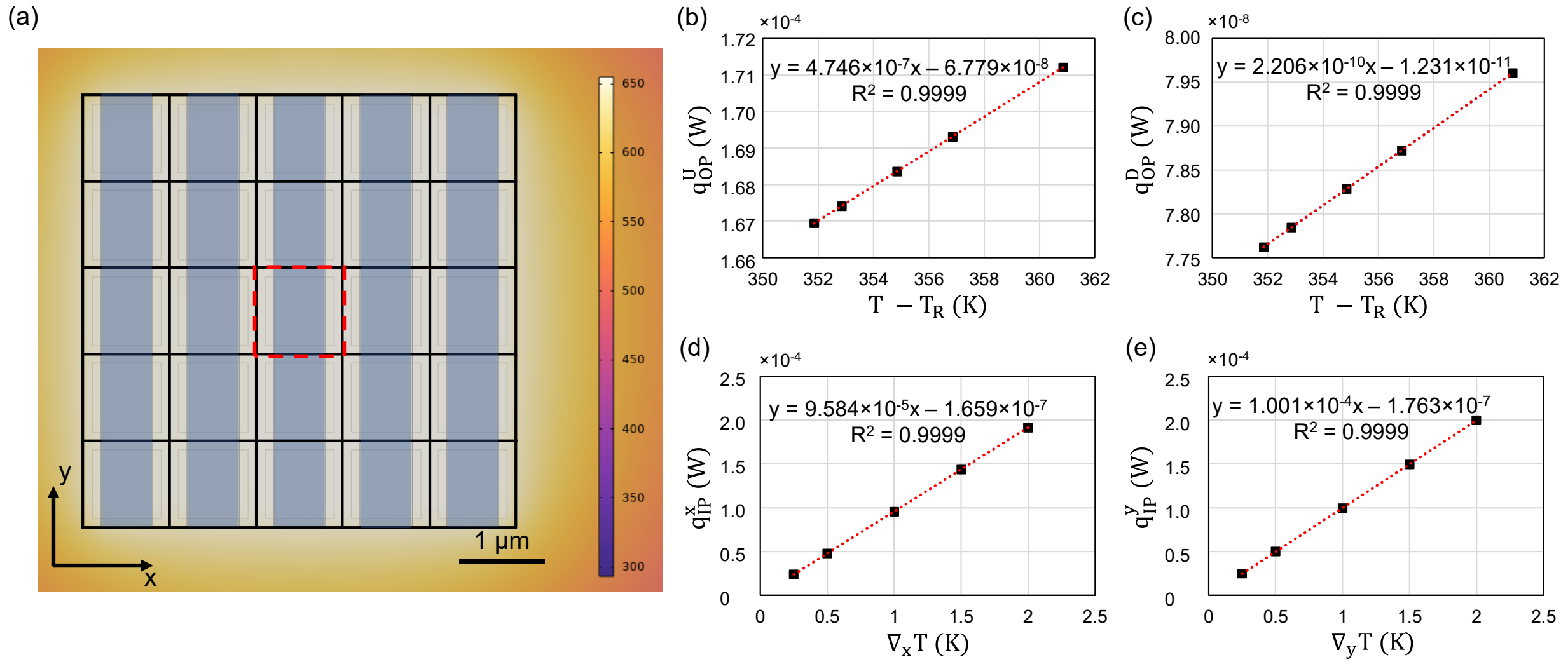

Fig. 2. (a) FEM simulation used to determine the thermal constitutive relations of the unit cells. (b-e) FEM simulated heat transfer rates along different directions (the symbols are defined in Fig. 1b as well as Eqs. 1 and 2).

To obtain the constitutive parameter $C_1$, we set all 25 unit cells to have the same temperature, i.e., with zero in-plane temperature gradient. The simulation allows us to validate Eq. (1), i.e., that the out-of-plane heat fluxes is linearly scales with the temperature difference $T$ - $T_R$ between the unit cell's temperature and room temperature. Figures 2(b) and (c) show the dependence of the heat transfer rates on the temperature difference $T$ - $T_R$, where the sum of the slopes of the two lines yields $C_1$. Because of the layer structure, heat dissipation primarily occurs through the substrate. Next, a linear temperature gradient across the simulation domain is defined by setting the temperatues of all 25 unit cells. The heat transfer rates along $x$- and $y$-directions are then computed (Figures 2(d) and (e)) to infer $C_2$. We also verify that the out-of-plane fluxes barely changes with varying in-plane temperature gradients, which corroboates our hypothesis that the heat flow into a unit cell can be decomposed into independent components in in-plane and out-of-plane directions, respectively. These results support our hypothesis that the thermal behavior of the unit cell can be characterized using two constitutive parameters $C_1$ and $C_2$ with adequate accuracy.

The conclusion above holds provided that the two assumptions we made hold: 1) the temperature within a unit cell is approximated as uniform; and 2) the in-plane temperature gradient is much smaller than the out-of-plane temperature gradient. The former assumption breaks down as the unit cell size increases. The latter assumption is in general valid in doped Si heaters, since the in-plane thermal conductance (given the exceptionally high thermally conductivity of Si) is far larger than that in the out-of-plane direction.

Under the steady-state condition, the net heat flow out of each unit cell must be exactly compensated by the heat generation. The constitutive relations Eqs. 1 and 2 therefore allow us to straightforwardly compute the heat source distribution $g(x, y)$ from $T(x, y)$. In the following, we show how to derive the doped filament width $W(x, y)$ from $g(x, y)$. Summing up the heat generation rate from all unit cells along a conducting filament of doped Si yields the total electric power for the entire filament since it is a pure resistive element:

$$\sum_{i}^{n} I^2 R(x_i, y_i) = IV \tag{3}$$

Eq. 3, coupled with the Joule's law of electric heating applied to each filament segment,

$$g(x_i, y_i) = I^2 R(x, y_i) \tag{4}$$

allows deterministic solution of all $R(x, y)$ and hence $W(x, y)$. Here $V$ is the applied voltage across the two heater contacts, $I$ is the current of passing through the filament, $g(x_i, y_i)$ denotes the heat generation rate of the $i^{th}$ unit cell segment in the filament, and $R(x_i, y_i)$ gives the segment's resistance.

The procedures above allows inverse design of heaters when the target temperature and in-plane temperature gradient (necessary to solve the in-plane heat transfer rate from Eq. 2) are known for each and every unit cell. However, typical heater inverse design problems do not specify the temperature gradient at the boundaries. To resolve this issue, an iterative optimization process is implemented as summarized in Figure 3. We define three zones in a heater: a target zone, a transistion zone, and an peripheral zone. The target zone is the actual domain within which the inverse design is performed; the peripheral zone is the area where the temperature drops considerably such that the second order derivative of the temperature profile at its outer boundary becomes negligible; and the transition zone has a smoothly varying temperature profile that bridges the two other zones. Initially, a temperature map $T_{target}$ is defined. The $T_{target}$ in the target zone is the design target, and the $T_{target}$ in the other two zones is some smooth function as an initial guess. Next, we perform inverse design across all three zones to obtain an initial heater design. An FEM simulation is then executed across the entire domain to generate a temperature map $T_{response}$. The $T_{response}$ in the peripheral zone is set as the new $T_{target}$ for the next iteration. The $T_{target}$ in the target zone remains the same in the subsequent iteration and an exponential function is used as the $T_{target}$ in the transition zone to produce a smooth transition. The process reiterates until the $T_{response}$ within the target zone meets the design specifications (more on this later). We found that the algorithm is robust and converges independent of the initial $T_{target}$ heuristic in the peripheral zone, as we will demonstrate in the next section. The process also converges to a design with satisfactory accuracy quite quickly: in the examples we give in the following section, the maximum number of iterations used is only 5. Therefore, even though this iterative optimization process still requires modeling of the entire heater device, the computational load is well manageable.

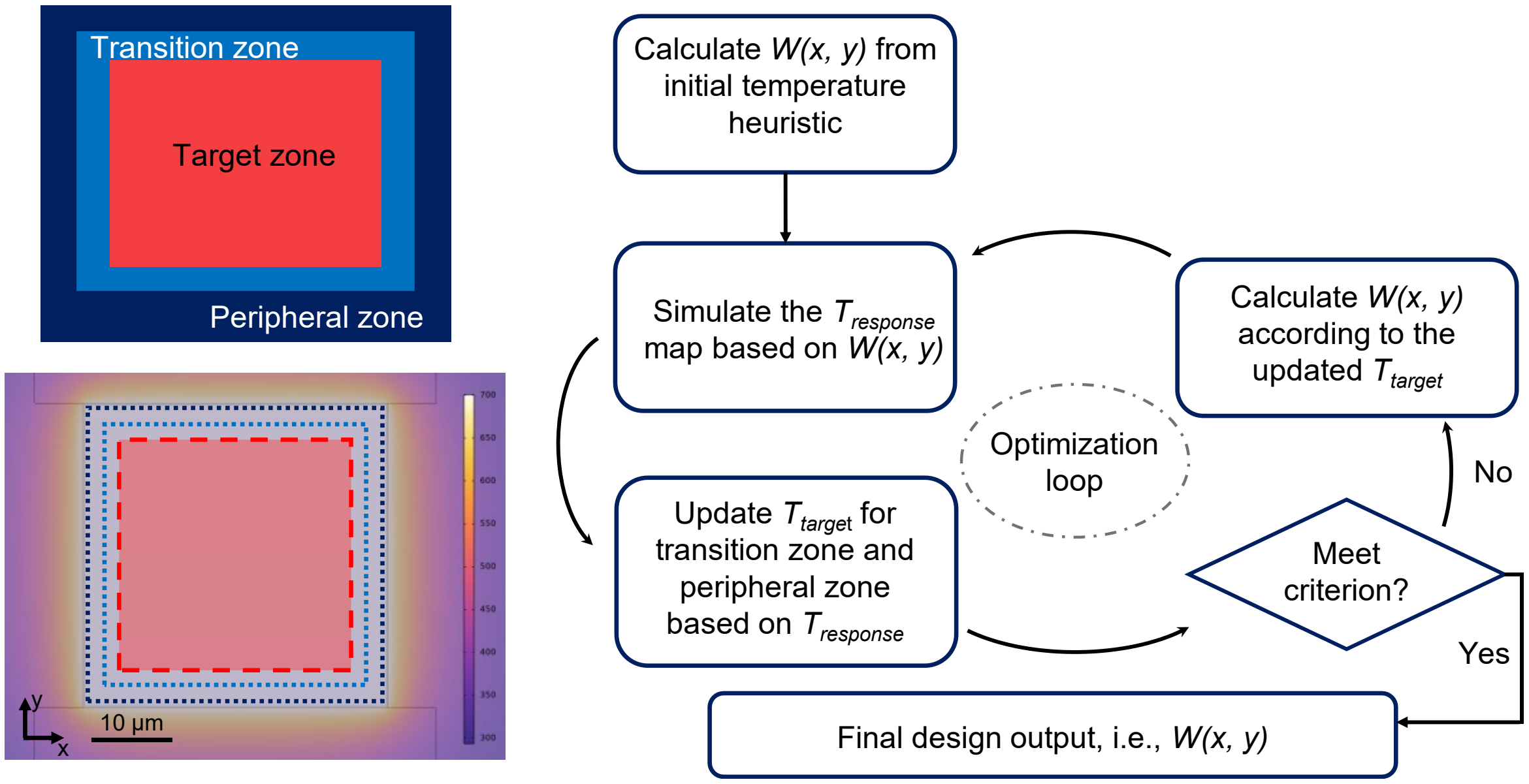

Fig. 3. Design process flow to determine the boundary condition: the top-left inset schematically illustrates the three zones, and the bottom-left inset marks the three zones used in the example in Fig. 4b.

## Result and Discussion

Here we present a set of heater design examples derived from our inverse-design method. The first example is a square doped Si heater targeting uniform temperature distributions. Figure 4 compares the temperature profile from a reference heater with a uniform doping concentration (Figures 4(a) and (c)) and that of an inverse-designed heater (Figures 4(b) and (d)). In both cases, the heater has an area of 38 × 38 μm$^2$. The target temperature profile is set to be 700 K across the center 30 × 30 μm$^2$ area. Figure 4(a) plots the temperature map of a reference heater with a uniform doping concentration, show a significant temperature change of 102 K from the center to edge within the 30 × 30 μm$^2$ target zone. In comparison, the inverse-designed heater exhibits a uniform temperature averaged at 699.7 K across the target zone with a root-mean-square (RMS) variation of only 0.18 K.

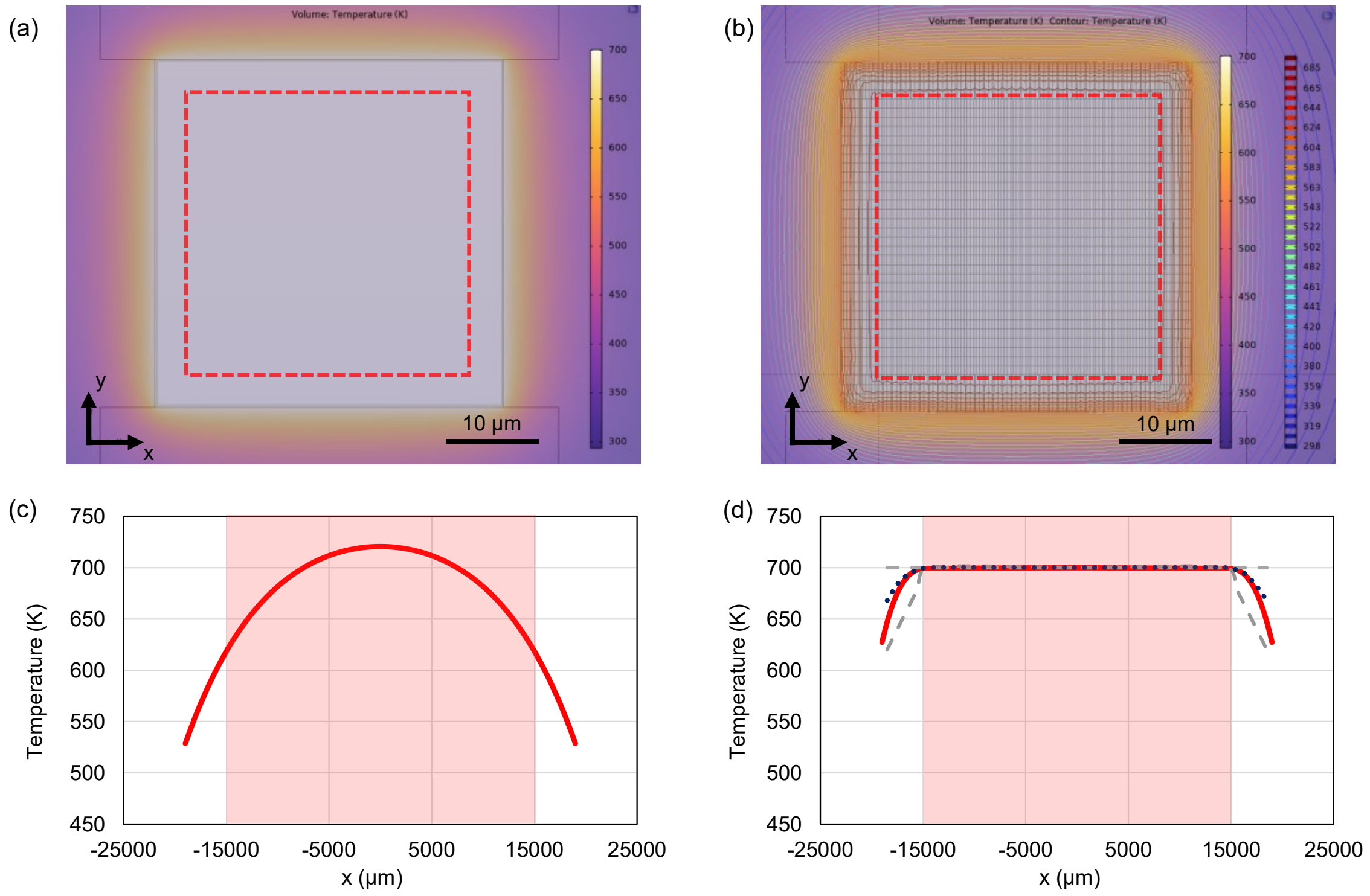

Fig 4. Temperature profiles of (a, c) a reference micro-heater with uniform doping and (b, d) an inverse-designed micro-heater targeting 700 K uniform temperature distribution. The red dotted lines encircle the target zone. (a) and (b) present in-plane temperature maps on the surfaces of both heaters. The solid red lines in (c) and (d) are 1-D sections of temperatures along the center planes of the heaters. The gray dash lines in (d) show two initial $T_{target}$ options tested in our design and both converged upon the same design. The blue dotted line gives the $T_{target}$ used in the final iteration of the optimization cycles.

In this example, our design figure-of-merit (FOM) adopted during the optimization process illustrated in Figure 3 is given as $1/[(T - T_{target})T_{std}]$, where $T - T_{target}$ is averaged over the target zone and $T_{std}$ is the standard deviation of temperature across the target zone. This FOM prioritizes temperature uniformity over the absolute temperature. As a result, the optimization produces a heater design with uniform 705.5 K temperature. This deviation from the initial 700 K target can be easily corrected by slightly decreasing the applied voltage to obtain the results shown in Figure 4(b) and (d). The optimization procedure is also largely agnostic to the choice of initial $T_{target}$. As an example, two overly simplistic $T_{target}$ options (gray dash lines in Figure 4(d)) were tested: one is a uniform 700 K temperature profile throughout all three zones and one assumes a linear temperature gradient in the transistion and peripheral zones that drops the temperature to ambient at the boundary. Both yield identical outcomes after the optimization procedure.

We further test uniform-temperature heater designs with different sizes of target zones, $10 \times 10$ μm$^2$ and $40 \times 40$ μm$^2$ (Figures 5(a) and (b)). It is worth pointing out that instead of executing the same optimization procedures in Figure 3, we directly adopted the final $T_{target}$ for the transition and peripheral zones from the prior heater design (the blue dot curve in Figure 4(d)). This important feature provides a facile means to scale heater designs while completely circumventing the need for full-heater-scale FEM simulations.

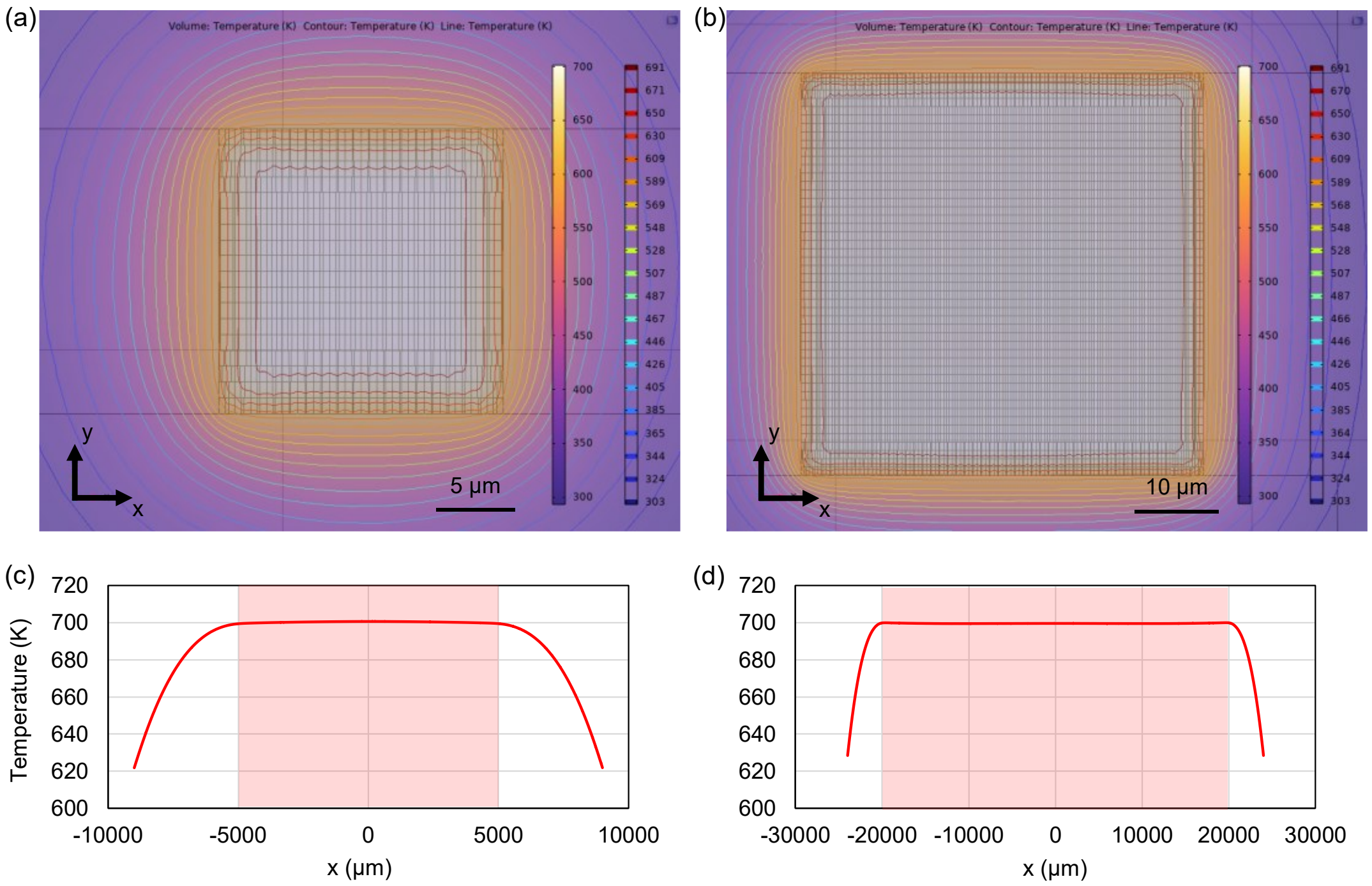

Fig 5. FEM-simulated (a, b) surface temperature distributions and (c, d) 1-D temperature sections of inverse-designed heaters with (a, c) 10 × 10 μm$^2$ and (b, d) 40 × 40 μm$^2$ target zones.

To demonstrate versatility of our method, we apply the inverse design scheme to implement two non-uniform temperature profiles. The first example seeks to achieve a linear gradient temperature along the diagonal direction of a square heater as shown in Figures 6(a) and 6(b). The target zone is 30 × 30 μm$^2$. While the temperature distribution agrees well with the design target near the center of the heater, some deviations are observed near its edges. We note that Figures 6(a) and 6(b) show optimization results that have converged, and therefore the remaining deviations are indicative that the initial temperature target is physically unattainable given the pre-defined boundary conditions. In this case, our inverse design scheme provides a facile means to assess if the design target is realistic.

The second example is a $T_{target}$ map that features areas with depressed/raised temperatures assuming the shape of an MIT logo within a 60 × 40 μm$^2$ target zone. Figures 6(c) and 6(d) display the FEM-simulated heater temperature distributions which closely match our design target. These examples indicate that our method can be broadly applied to produce heater designs with on-demand temperature distributions [10] provided that the design target is physically viable.

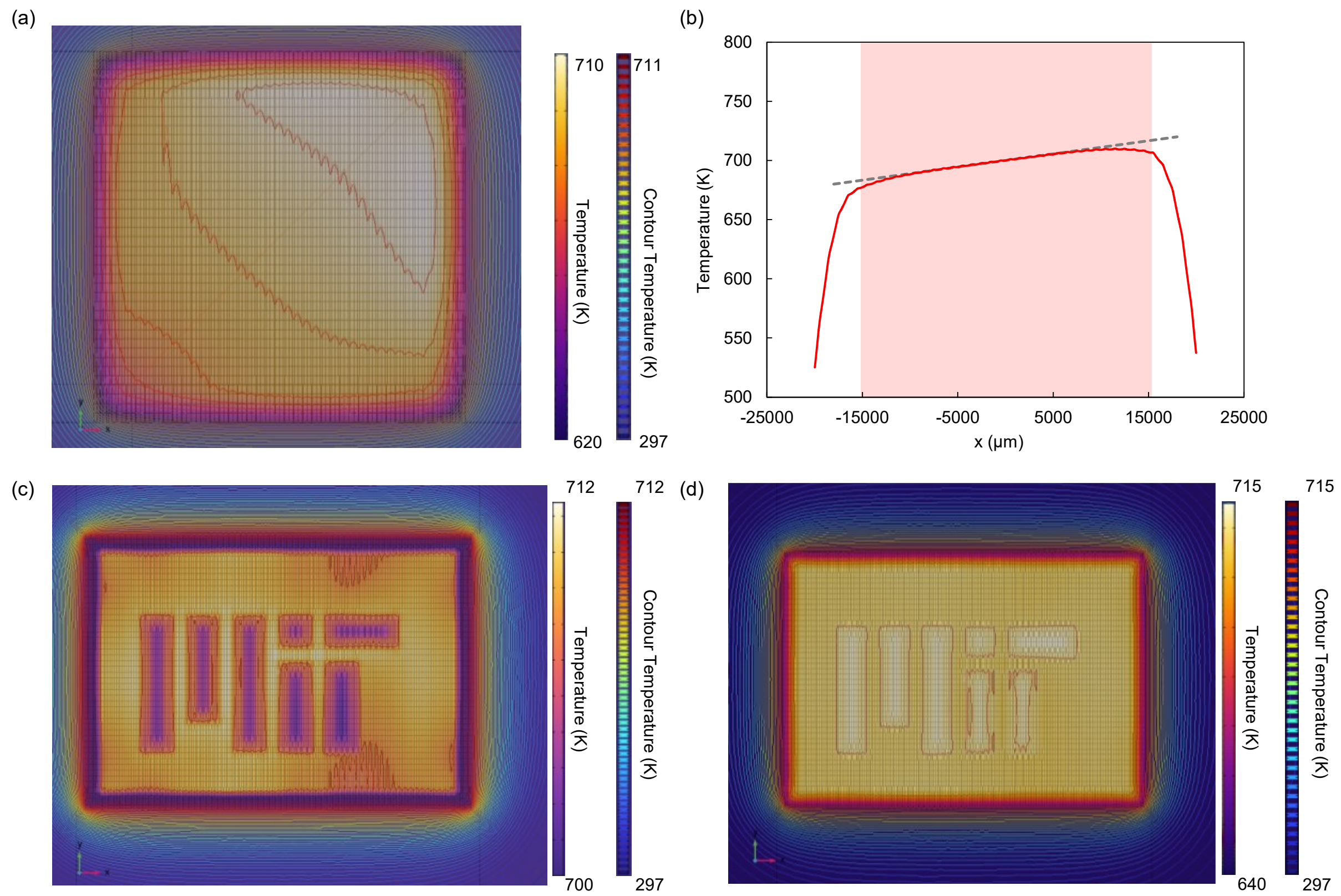

Fig 6. FEM-simulated surface temperature distributions of inverse-designed heaters: (a) a square heater with a linearly varying temperature profile along its diagonal and (b) its temperature distribution long the diagonal line; (c, d) heaters designed to produce an MIT logo in the form of areas with (c) depressed and (d) raised temperatures.

The method is not limited to planar heater structures and can be applied to heaters with uneven surface morphologies as well. As an example, we consider a doped Si heater incorporating a ridge waveguide across its center as shown in Figure 7. Similar heater configurations are commonly adopted in photonics, for instance to switch the structure of phase change materials (PCMs) [15], where the ability to precisely engineer temperature profiles is essential for controlling the materials' phase transition kinetics and avoiding damage due to hot spots [17]. In this example, we assumed a ridge waveguide structure fabricated in a 220 nm thick silicon-on-insulator layer with an etch depth of 110 nm and a ridge width of 450 nm. The waveguide is cover by a 100-nm thick silicon dioxide cladding (Figure 7(c)). Two sets of constitutive parameters were computed corresponding to two types of unit cells: unit cells on the waveguide and unit cells outside the waveguide. With a design target of uniform temperature distribution at 700 K, the inverse design scheme leads to an average temperature of 700.3 K with a standard deviation of 0.16 K, as shown in Figure 7(a) where the waveguide is highlighted by the white solid line and the target zone is within the red dash line.

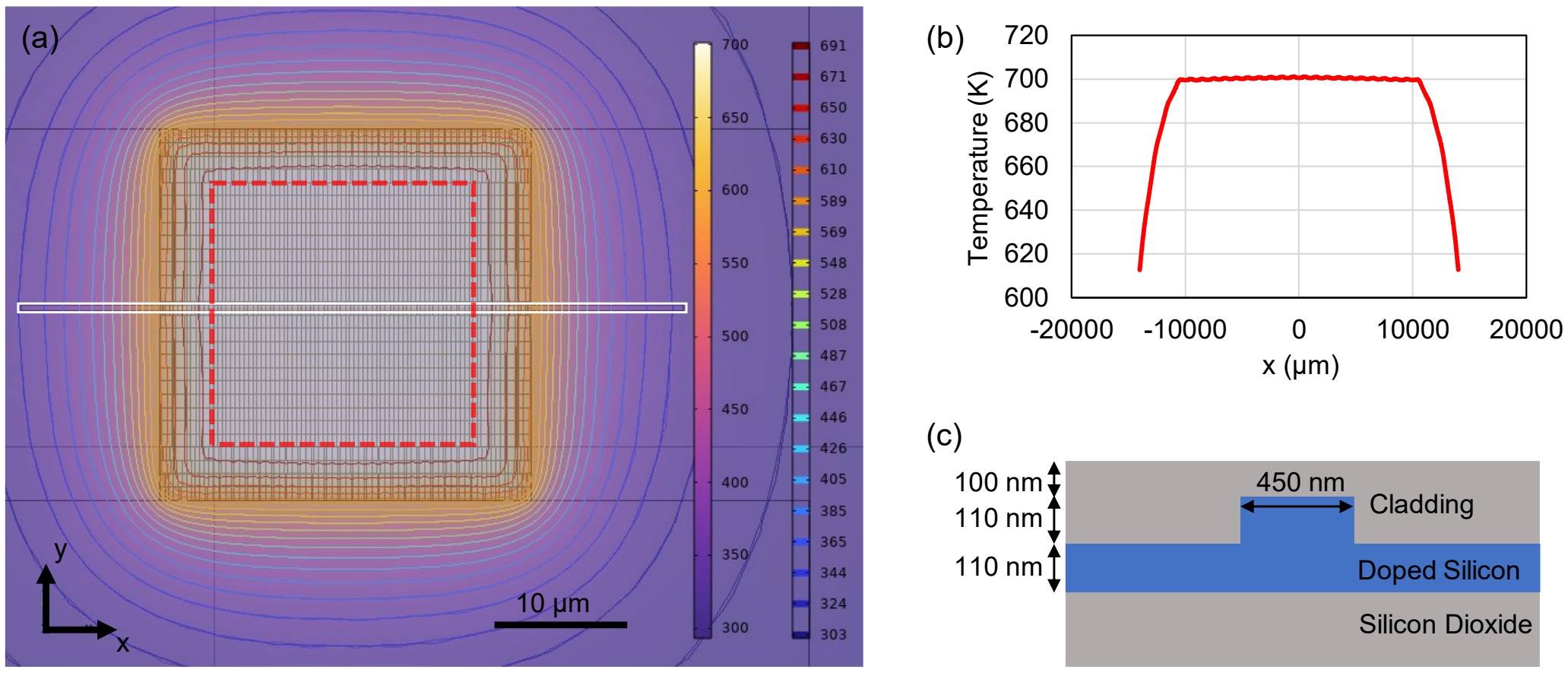

Fig 7. (a) FEM-simulated surface temperature distributions of an inverse-designed waveguide heater; (b) temperature profile along the waveguide; (c) the waveguide cross-sectional structure.

The examples we cited here all assume heaters with a rectangular shape. Our approach, however, can be readily extended to heaters with irregular geometries. In the latter case the unit cells will no longer be rectangular but rather parallelogram in shape and might not be uniform in size. Thermal constitutive properties for each size and shape of unit cells need to be calibrated following protocols similar to that in Figure 2 and compiled into a lookup table. They can then be used in conjunction with Eqs. (1) and (2) to determine the heat generation map and the subsequent design process flow is similar.

The accuracy of the inverse-design approach is presently limited by mesh size of the FEM simulations. Since the FEM model has a minimum mesh size which all structures must snap to, this inevitably results in deviation of the local filament width from the analytically derived values from Eqs. (3) and (4). For the design illustrated in Figures 4, 5, 6 and 7, the minimum mesh size (constrained by the computer memory capacity) is approximately 5 nm. This is comparable to, and in some cases considerably larger than the typical filament width variations between neighboring unit cells. This limitation partially accounts for the observed deviations from design targets in our examples.

## Conclusion

In conclusion, we proposed and validated a generic inverse-design framework to engineer doped Si micro-heaters' temperature profiles. The approach can be fully automated with minimal-to-none human intervention. Moreover, it circumvents the need for extensive full heater simulations and instead uses the thermal constitutive properties of small unit cells to construct the heater design. While we demonstrated the inverse design approach in doped Si heaters, the concept can be adapted to other resistive heater systems as well. Given the ubiquity of resistive heaters in microsystems, the ability to generate on-demand temperature distributions foresee many exciting potential applications in photonics, electronics, and micro-mechanics.

## ACKNOWLEDGEMENTS

The funding support is provided by the National Science Foundation under awards 2329088 and 2225968. We cordially acknowledge Dr. Kiumars Aryana and Dr. Yifei Zhang for insightful discussions and technical assistance.

## AUTHOR DECLARATIONS

**Conflict of interest**

The authors have no conflicts to disclose.

**DATA AVAILABILITY**

The data that support the findings of this study are available from the corresponding author upon reasonable request.